\documentclass{article}

\usepackage{arxiv}

\usepackage[utf8]{inputenc} 
\usepackage[T1]{fontenc}    
\usepackage{hyperref}       
\usepackage{url}            
\usepackage{booktabs}       
\usepackage{amsfonts}       
\usepackage{nicefrac}       
\usepackage{microtype}      
\usepackage{lipsum}		
\usepackage{graphicx}
\usepackage{doi}

\usepackage[utf8]{inputenc}
\usepackage{subcaption,color, graphics}
\usepackage{enumerate}
\usepackage{soul} 
\usepackage{multirow}
\usepackage{comment, amsmath}
\newcommand{\ignore}[1]{}
\newtheorem{insight}{{\bf Insight}}
\usepackage{natbib}
\usepackage{graphicx}
\UseRawInputEncoding

\begin{document}

\title{Internet-based Social Engineering Attacks, Defenses and Psychology: A Survey}

\author{Theodore Longtchi\textsuperscript{1, *}\\
\and
{\textbf{Rosana Monta\~nez Rodriguez}}\textsuperscript{2, *}\\
\and
Laith Al-Shawaf\textsuperscript{3}\\
\and
Adham Atyabi\textsuperscript{1}\\
\and
Shouhuai Xu\textsuperscript{1}\\
}

\maketitle

\bigskip
{\textbf{1}} Department of Computer Science, University of Colorado Colorado Springs, Colorado Springs, CO 80918
\\
{\textbf{2}} Department of Computer Science, University of Texas at San Antonio, San Antonio, TX 78249 
\\
{\textbf{3}} Department of Psychology, University of Colorado Colorado Springs,	Colorado Springs, CO 80918 
\bigskip
\newline
* Shared first-author

\begin{abstract}
Social engineering attacks are a major cyber threat because they often serve as a first step for an attacker to break into an otherwise well-defended network, steal victims' credentials, and cause financial losses. The problem has received due amount of attention with many publications proposing defenses against them. Despite this, the situation has not improved. In this paper, we aim to understand and explain this phenomenon by looking into the root cause of the problem. To this end, we examine the literature on attacks and defenses through a unique lens we propose --- {\em psychological factors (PFs) and techniques (PTs)}. We find that there is a big discrepancy between attacks and defenses: Attacks have deliberately exploited PFs by leveraging PTs, but defenses rarely take either of these into consideration,
preferring technical solutions. 
This explains why existing defenses have achieved limited success. This prompts us to propose a roadmap for a more systematic approach towards designing effective defenses against social engineering attacks.
\end{abstract}


\ignore{

}
\keywords{Social engineering attacks, email-based attacks, website-based attacks, online social network-based attacks, psychological factors, psychological techniques, phishing, deception}

\thanks{Correspondence: {\tt sxu@uccs.edu}}

\section{Introduction}

It is often said that humans are the weakest link in cybersecurity. However, the cause of this phenomenon is poorly understood, and solutions are elusive. These issues motivate us to take a deeper look into the following problems: (i) What is the root cause that enables social engineering attacks? (ii) Why have existing defenses achieved very limited success in mitigating social engineering attacks? (iii) What kinds of research need to be done in order to adequately mitigate social engineering attacks? 
In order to answer these questions, we need to narrow down the scope, since human factors are such a broad topic (see, e.g., \cite{XuBookChapterSocialEngineering2022}). This prompts us to focus on Internet-based social engineering attacks to which humans often fall victim. Specifically, we will focus on three classes of social engineering attacks:  email-based, website-based, and online social network (OSN)-based social engineering attacks.

The importance of the aforementioned motivating problems in the context of Internet-based social engineering attacks is demonstrated by the multitude of studies on the aforementioned attacks and defenses against them, and by the many existing surveys from various perspectives of the problem (cf. 
\cite{
das2019sok, 
khonji2013phishing, zaimi2020survey, dou2017systematization, 
almomani2013survey, 
basit2020comprehensive, ali2020survey, jain2021survey, sahu2014survey, da2020heuristic, alabdan2020phishing, vijayalakshmi2020web, jampen2020don, chanti2020classification, rastenis2020mail, gupta2018defending, chiew2018survey, aleroud2017phishing, yasin2019contemplating, montanez2020human, gupta2016literature, heartfield2015taxonomy, alharthi2020taxonomy, salahdine2019social} and the extensive list of references therein). It is also shown by the scale of financial losses incurred by these attacks; for example,
FBI reports a \$26B loss
between June 2017 and July 2019 associated with attack emails that contain instructions on approving payments to attackers and pretend to come from executives
 \cite{fbi_2020}. 
However, our examination shows that prior studies focus on technical solutions which mainly leverage Artificial Intelligence/Machine Learning (AI/ML) to design various kinds of defense systems. This may be understood as follows: Since these attacks often leverage information technology as a means to wage attacks (e.g., many attacks do not have counterparts in the era prior to the emergence of cyberspace), it has been largely treated as a technological problem. This is evidenced by the fact that most existing defenses against them leverage AI/ML techniques. As a result, very few studies take a close look at the root causes of social engineering techniques; the present study fills this void.

\smallskip

\noindent{\bf Our Contributions}. In this paper we systematize Internet-based social engineering attacks and defenses through the lens of psychology. We resolve the aforementioned motivating problems with a systematic characterization of attacks and defenses with respect to psychological factors and techniques that contribute to human susceptibility to social engineering attacks.
Specifically, we make the following contributions.

First, we systematize the human {\em psychological factors} (PFs) that have been deliberately exploited by attackers to wage effective attacks, while noting that these factors have been scattered in the literature. The success of these attacks suggests that attackers have to give the problem due diligence in understanding PFs, especially those factors that can be exploited effectively. In order to deepen our understanding of how these PFs can be exploited to wage attacks, we further systematize what we call {\em psychological techniques} (PTs), which can be seen as the means of exploiting PFs. The PFs and PTs help understand the root cause that enable social engineering attacks from a psychological perspective. To the best of our knowledge, this is the first systematization of PFs and PTs with respect to social engineering attacks.

Second, we systematize Internet-based social engineering attacks, with emphasis on the PFs they exploit. This is made possible by the ``bridge'' of PTs. Moreover, we systematize defenses with an emphasis on whether a defense leverages certain PFs. We find that very few do. This means that most defenses are designed without considering the root cause of these attacks, which explains why current defenses have achieved limited success. 

Third, the above finding prompts us to propose a research roadmap towards designing effective defenses. The roadmap is centered at creating a psychological framework tailored to social engineering attacks. The framework aims to tailor psychological principles to the cyber domain.
The framework highlights the role and relevance of each PF, including the relationships between each. These relationships are important because they are not necessarily independent of, or orthogonal to, each other. The envisioned quantitative understanding will guide us to design effective defenses in the future.

\smallskip


\noindent{\bf Related Work}.
We focus on social engineering attacks in cyberspace, which is in contrast to social engineering attacks in the physical world \cite{XuBookChapterSocialEngineering2022}. 
Table \ref{table:1} contrasts the present survey with related previous surveys through the coverage of the following attributes: PFs
are the human attributes that can be exploited by social engineering attacks (e.g., greed); PTs describe how social engineering attacks exploit PFs;
Attacks 
waged by social engineering attackers (e.g., whaling); Defenses 
which have been proposed in the literature.
For example, 
Khonji et al. \cite{khonji2013phishing} surveyed phishing definitions and detection methods. When compared with these studies, we stress two fundamentally important aspects: PFs and PTs, because we must understand them before we can design effective defenses. Indeed, this perspective has three immediate payoffs as shown in the paper: (i) we can map social engineering attacks to PFs through the ``bridge'' of PTs; (ii) defenders largely lag behind attackers because most defenses do not adequately take PFs into consideration, even though attacks have been regularly exploiting PFs in crafty ways, explaining the limited success of current defenses; (iii) this understanding prompts us to propose a research roadmap towards the development of effective defenses against social engineering attacks.

\begin{table}[!htbp]
\centering
\begin{tabular}{|r|c|c|c|c|c| }
 \hline
Ref. & PFs & PTs & Attacks & Defenses & Relationships \\
  \hline
 \cite{das2019sok} &$\surd$ (6) &  & $\surd$ & $\surd$ &  \\ \hline
 \cite{khonji2013phishing} & & & $\surd$ & $\surd$ & \\ \hline
 \cite{zaimi2020survey} & & & $\surd$ & $\surd$ &  \\ \hline
 \cite{dou2017systematization} & & & $\surd$ &$\surd$ &   \\ \hline 
 \cite{almomani2013survey} & & & $\surd$ & $\surd$ &  \\ \hline 
 \cite{basit2020comprehensive} & & & $\surd$ & $\surd$ &  \\ \hline 
 \cite{ali2020survey} & & & & $\surd$ & \\ \hline 
 \cite{jain2021survey} & & & $\surd$ & $\surd$ &  \\ \hline 
 \cite{sahu2014survey} & & & $\surd$ & $\surd$ & \\ \hline 
 \cite{da2020heuristic} &  &  & $\surd$ & $\surd$ &  \\ \hline 
 \cite{alabdan2020phishing} & $\surd$ (19) &  & $\surd$ & $\surd$ & \\ \hline 
 \cite{vijayalakshmi2020web} & & & & $\surd$ &  \\ \hline 
 \cite{jampen2020don} &$\surd$ (18) &  & & $\surd$ &  \\ \hline 
 \cite{chanti2020classification} & & & $\surd$ & $\surd$ &   \\ \hline
 \cite{rastenis2020mail} & & & $\surd$ & &  \\ \hline
 \cite{gupta2018defending} & & & $\surd$ & $\surd$ &  \\ \hline
 \cite{chiew2018survey} & & & $\surd$ & & \\ \hline
 \cite{aleroud2017phishing} & & & $\surd$ & $\surd$ &  \\  \hline
 \cite{yasin2019contemplating} & & & $\surd$ & &  \\ \hline 
 \cite{montanez2020human} & $\surd$ (19) &  & & $\surd$ &  \\ \hline
 \cite{gupta2016literature} & & & $\surd$ & $\surd$ &  \\ \hline 
 \cite{heartfield2015taxonomy} & $\surd$ (2) &  & $\surd$ & $\surd$ &  \\ \hline 
 \cite{alharthi2020taxonomy} & & & $\surd$ & $\surd$ &  \\ \hline
 \cite{salahdine2019social} & & & $\surd$ & $\surd$ &  \\ \hline
 The present paper & $\surd$ (41) & $\surd$ (13) & $\surd$ & $\surd$ & $\surd$  \\ \hline
\end{tabular}
\caption{Comparison between existing surveys and ours, where a number in parentheses is the number of PFs or PTs discussed in a paper. Only \cite{montanez2020human} and the present paper systematize PFs even though \cite{montanez2020human} only discusses 19 PFs (which are significantly less comprehensive than the 41 discussed in the present paper) ; the others merely mention some PFs. Only the present paper explores the relationships between the PFs, PTs, attacks, and defenses.
}
\label{table:1}
\end{table}

\noindent{\bf Paper Outline}. 
Section \ref{sec:methodology} reviews some preliminary psychology knowledge and describes our methodology. 
Section \ref{sec:overview} presents an overview of our study following the methodology.
Section \ref{sec:vulnerabilities}
systematizes PFs and PTs. Section \ref{sec:attacks} systematizes social engineering attacks. Section \ref{sec:defenses} systematizes defenses against social engineering attacks.  Section \ref{sec:sok} systematizes the relationships between PFs, PTs, social engineering attacks, and social engineering defenses. Section \ref{sec:research-directions} presents a roadmap for future research directions. Section \ref{sec:conclusion} concludes the present study. 

\section{Psychological Preliminaries and Study Methodology} 
\label{sec:methodology}


\subsection{Psychological Background Knowledge}


We briefly review the psychological background knowledge that is helpful for understanding the paper.
This knowledge serves as a baseline for guiding us in defining what constitute as PFs.

\smallskip

\noindent{\bf Big Five Personality Traits (BFPT)}. The Five Factor Model of personality traits -- also known as the Big Five -- refers to the five factors that constitute the basic structure of human personality \cite{goldberg1981language}. These factors, discovered using factor analysis and other statistical techniques, are {\sc openness}, {\sc conscientiousness}, {\sc extraversion}, {\sc agreeableness}, and {\sc neuroticism}. The evidence for the existence and robustness of the Big Five comes from numerous studies conducted in different languages and across cultures over the span of many decades \cite{costa2008revised, digman1990personality, mccrae1992introduction}. These basic personality traits are relatively stable across the lifespan, and they predict life outcomes ranging from career success to likelihood of divorce to lifespan longevity \cite{soto2019replicable}. 
They even appear to be present in other species \cite{nettle2006evolution}. Given this robustness and for the purposes of the present paper, each of the five factors constitutes a PF.

\ignore{\color{ForestGreen}These five traits refer to {\sc Openness}, {\sc Conscientiousness}, {\sc Extraversion}, {\sc Agreeableness}, and {\sc Neuroticism}. They were discovered using cross-cultural studies and factor analysis. They describe the basic structure of human personality.

\footnote{I do not think this is correct because BFPT was not originally introduced for the purpose of coping with social engineering attacks. Laith: need your expertise} 
BFPT describes individual differences in terms of characteristic thoughts, feelings and behaviours \cite{frauenstein2020susceptibility}. It turns out that these personality traits are also factors that affect individuals' susceptibility to social engineering attacks \cite{alohali2018identifying}.} 

\smallskip

\noindent{\bf Cialdini's Principles of Persuasion}. Persuasion Principles are a set of strategies used to influence individuals into behaving in a desired way. These principles, derived from field studies on sales and marketing \cite{cialdini2009influence}, include: (i) {\sc liking} which denotes being easily influenced by those one likes or those with common beliefs as them; (ii) {\sc reciprocation} which denotes feeling obliged to return a favor; (iii) {\sc social proof} (conformity) which denotes imitating the behaviours of others; (iv) {\sc consistency} (commitment) which denotes consistency of behaviour or sticking to a promise; (v) {\sc authority}, which denotes submitting to experts or obeying orders from one's superior or authoritative figures;
and (vi) {\sc scarcity} which denotes placing more value on things that are in short supply. When persuasion is undetected, it encourages the use of heuristic reasoning \cite{cialdini2009influence, cialdini1998social}; when persuasion is detected, it results in a negative response towards the message \cite{kirmani2007vigilant}. The use of persuasion in social engineering messages has been studied extensively \cite{lin2019acm, rajivan2018creative, ferreira2015analysis, stajano2011understanding, van2019cognitive}.
For the purposes of this paper, each of the six persuasion principles constitutes a PF. 

\subsection{Study Methodology}

\ignore{

Figure \ref{fig:Methodology} highlights our systematization methodology, which can be understood as follows.

\begin{figure}[htbp!]
\centering
\includegraphics[scale=.67]{images/Methodology_compact.JPG}
\caption{Schematic representation of the methodology}
\label{fig:Methodology}
\end{figure}

}

\noindent{\bf Scope}. 
Since social engineering attacks are a broad topic, we choose to focus on Internet-based social engineering attacks, especially the ones exploiting emails, websites, and online social networks (OSNs). Therefore, the term ``social engineering attacks'' or simply ``attacks'' in this paper refers to these attacks. We will use terms ``individuals'' and ``users'' interchangeably; the term ``victims'' refer to the users that are successfully compromised by social engineering attacks.

\smallskip

\noindent{\bf Methodology}. 
In order to understand why humans are susceptible to social engineering attacks, we aim to systematize the PFs and PTs that have been, or could be, exploited by attacks. In this paper, the term {\em psychological factor} is used to represent the psychological attributes that can be exploited by attacks (i.e., {\em what to exploit}). 
In order to make our definitions of PFs psychological sound, we leverage the afore-reviewed BFPT and Principles of Persuasion among others from Cognitive Psychology, to define PFs to guide us in identifying other psychological attributes that can be deemed as PFs in a psychological sense.
By contrast, the term {\em PT} is used to describe {\em how attacks exploit these factors}. This distinction turns out to be useful because the PTs will be leveraged to build a ``bridge" for mapping attacks to PFs. In other words, PTs build a bridge between the two disciplines of psychology and cybersecurity. Note that one PT can exploit multiple PFs and one PF can be exploited by multiple PTs.

\begin{figure}[!htbp]
\centering
\includegraphics[width=.8\textwidth]{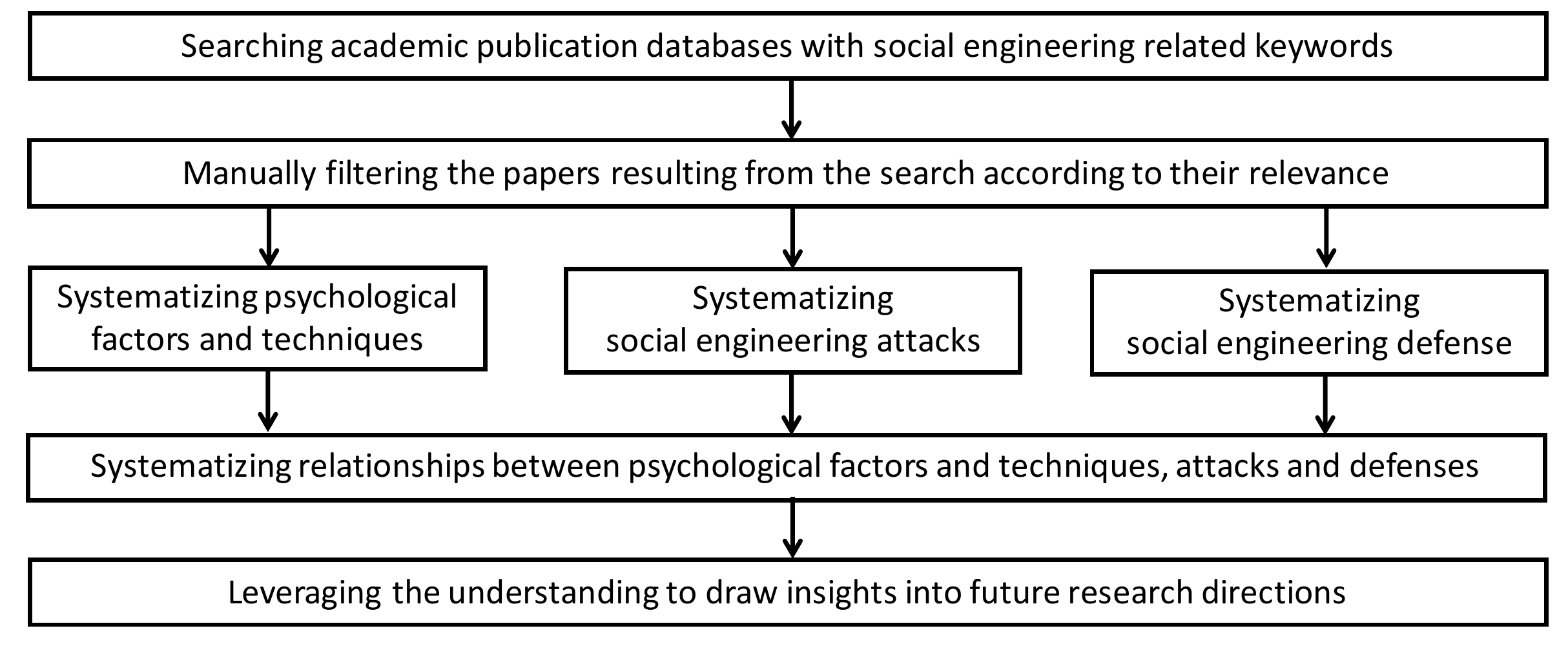}
\caption{Methodology of our study
}
\label{fig:methodology}
\end{figure}

Figure \ref{fig:methodology} highlights our methodology, which can be adopted or adapted by others to conduct their own studies. The methodology consists of the following five steps. 
First, identify the relevant literature, ideally automatically or semi-automatically because there are so many publications. 
Second, manually filter the papers resulting from the previous step according to their relevance to the purpose of the study.
The preceding two steps correspond to {\em preparation}.
Third, extracting and systematizing the PFs, PTs, attacks, and defenses that are discussed in the papers resulting from the previous step.
Fourth, systematizing the relationships between the PFs, PTs, attacks, and defenses identified from the previous step. Fifth, leveraging the understanding resulting from the systematization to draw insights into future research directions.

Note that the methodology can have useful variants. For example, it can be adapted to search the literature in an iterative fashion when one does not know enough information about PFs, PTs, attacks, or defenses; in this case, it may be helpful to start the search with few well-known keywords, filtering the resulting literature, and extracting other PFs, PTs, attacks, or defenses for further search. This iteration can also be leveraged to assure whether some relevant papers are overlooked. As another example, one can incorporate the PFs, PTs, attacks, and defenses that are known to the investigator but are not found in the academic literature. As we will see, this is particularly relevant to attacks because there can be attacks that are discussed by practitioners in social media but are not investigated in the academic literature yet.

\section{Systematization Overview}
\label{sec:overview}

The overview corresponds to the five steps of the methodology.
First, to identify papers for manual filtering, we consider the following digital libraries: 
IEEE (including IEEE Symposium on Security and Privacy and IEEE Transactions), ACM (including ACM Conference on Computer and Communications Security and ACM Transactions), Usenix (including Usenix Security Symposium), and Network and Distributed System Security Symposium (NDSS), Elsevier, Springer, PlosOne, Wiley, Frontiers in Psychology, and Information \& Computer Security (ICS). These venues are selected because they are the main security venues or psychological venues that publish social engineering related papers.
We start the search with the only keyword ``social engineering'',  which yields thousands of papers. Since it is infeasible to manually filter this many papers, we add another keyword ``phishing'' (i.e., a paper is relevant when containing these two keywords simultaneously).
We choose phishing  because it is the most common social engineering attack \cite{touchstonesecurity-2021, kaspersky-2022} and may have been more widely studied from a psychological perspective. 

Second, the search identifies 663 papers. We manually examine these papers based on their treatment and relevance to the motivation and scope of the present study, leading to 154
papers which include 
24 
survey/review papers (while noting that the other references are cited for further exploration purposes, such as those published in psychology literature).  
We eliminate the papers that just mention social engineering and/or phishing without presenting substantial investigation; 268 papers fall into this category.
We further eliminate the ones that do not consider specific PFs or do not consider 
Internet-based social engineering attacks; 146 papers fall into this category.
Finally, we look into the remaining 249 papers and eliminate the ones that do not appear to present a high-quality exploration (which is a judgement call), ending up with the aforementioned 154 papers for systematization.

Third, given the 154 papers (including the 24 surveys),
we initially identified 63 PFs from these papers where each PF is discussed in one or multiple papers as a psychological factor that has an impact on social engineering attacks. We found that these PFs contain some redundant ones. This prompts us to eliminate the redundant ones as follows.
(i) If two PFs are considered similar to each other in a psychological sense, we keep the PF that is investigated in a quantitative fashion, representing a deeper understanding. (ii) If two PFs are considered redundant in a psychological sense but none of them is investigated in a quantitative fashion, we keep the PF that is most relevance to this paper.
(iii) If two PFs are considered redundant and both are investigated in a quantitative fashion, we keep the one that is more often used in the literature according to our psychological knowledge. This pre-processing leads to 41 PFs, 
which are PFs that are widely accepted in the psychology community, as reviewed above.
One example of the preceding (i) is the pair of PFs known as "false consensus effect" and "social proof", which are considered redundant in their psychological meanings. We keep the latter because it is investigated in a quantitative fashion \cite{das2014increasing}.
On example of the preceding (iii) is the pair of PFs known as "inattentiveness" and "lack of vigilance", which are redundant because they essentially represent the same psychological factor. Since they both have been investigated in a quantitative fashion \cite{tu2019users}, we keep the former because the former is used in the literature more often (perhaps because of its succinctness). 
After finalizing the 41 PFs, we leverage our psychological expertise to classify them into five classes: {\em cognitive},  {\em emotion}, {\em social psychology}, {\em personality and individual difference}, and {\em workplace} PFs.
These PFs help understand the root cause of social engineering attacks from a psychological perspective. 

In addition, we systematize the 13 PTs that have been mentioned in the literature. At this point, it is not clear to us how to further group these 13 PTs into a smaller number of categories. We hope future studies can resolve this issue.
We systematize attacks with respect to those PFs, including the PF(s) exploited by an attack and the extent to which a PF's impact on human susceptibility has been quantified.
We systematize defenses with respect to the attacks, leading to defenses against email-based, website-based, and OSN-based attacks, respectively. 


Fourth, we systematize the relationships between the PFs, PTs, attacks and defenses by proposing mappings between them. 
This is made possible because the PTs serve as a ``bridge'' between the attacks and the PFs. It is ironic to observe that only one PF has been leveraged by defenses. Nevertheless, the mapping gives a succinct representation of the state-of-the-art knowledge in this domain, and also allows us to understand the motivating problem of the cause of the limited success of existing defenses. 

Fifth, we propose a roadmap for future research to pursue the motivating problem of finding the solution to internet-based social engineering attacks.
The roadmap is prompted by the observation that very few empirical studies have been carried out to quantify the impact of PFs on the susceptibility of humans to social engineering attacks. The few studies that attempted quantitative studies have some drawbacks, such as involving a small number of participants (e.g., 53 participants \cite{welk2015will}) or participants being undergraduate students \cite{welk2015will,hong2013keeping,tu2019users} (rather than average users). 
The lack of quantitative results published in the literature (e.g., the impact of PFs) prevents us from conducting any meta-analysis.

\section{Systematizing Psychological Factors and Techniques}
\label{sec:vulnerabilities}

\subsection{Systematizing Psychological Factors (PFs)}

We categorize the PFs that may be exploited by Internet-based social engineering attacks into five groups, with 42 factors in total: (i) {\em cognitive PFs}, which describe how individuals process information; (ii) {\em emotion PFs}, which describe individuals' feelings, motivational state, and approaches or avoidance behaviors; (iii) {\em social psychology PFs}, which describe individuals' interpersonal attributes in various groups; (iv) {\em personality and individual difference PFs}, which are individuals' relatively stable attributes; and (v) {\em workplace PFs}, which describe cultural and organizational interactions within a workplace.  
Note that there is more than one way to divide these factors. We have adopted this framework because it is in line with the traditional branches and subdivisions in psychology, and because it may help readers decide where to direct their efforts in training victims to become less vulnerable -- but it is not a conclusive categorization.

It is worth mentioning that the categorization is somewhat subjective because several PFs can fall into more than one category. For example, we have listed {\sc overconfidence} as a {\em cognitive PF}, but it could reasonably be considered an {\em personality and individual difference PF} because there are stable individual differences in this factor. 

\smallskip

\noindent{\bf Cognitive PFs}.
These PFs describe how an individual processes information, including heuristics they may use, the knowledge they may possess, the confidence they may exhibit, and the attention they may give.
\begin{enumerate}
\item {\sc Cognitive Miser}. This PF describes one's use of decision-making heuristics, namely the use of mental shortcuts in a decision-making process \cite{mcalaney2020cybersecurity}. Generally speaking, people tend to be cognitive misers and rely more on heuristic-based processing to make decisions \cite{kim2021towards}. McAlaney and Hills \cite{mcalaney2020understanding} argued that people are motivated tacticians and will apply a cognitive miser (or naïve scientist) approach based on the urgency, perceived importance and complexity of the situation. 
\item {\sc Expertise}. This PF describes one's knowledge about a particular domain. 
Albladi and Weir \cite{albladi2020predicting} showed that {\sc expertise} plays a role in raising an individual's perception of risk associated with online social networks, but the perceived risk does not significantly increase individuals' competence in coping with these threats. Qualitatively speaking, expertise does not necessarily make one less vulnerable to social engineering attacks \cite{ghafir2016social, das2019all}; quantitatively speaking, expertise, when effectiveness, does make one more capable in coping with social engineering attacks (i.e., incurring lower false-positive and false-negative rates \cite{henshel2015trust}.  Redmiles et al. \cite{redmiles2020comprehensive} show that the {\sc expertise} associated with a given social-demographic background may affect the prioritization of advice in coping with online threats.
\item {\sc Overconfidence}. This PF describes individuals' tendency in having too much confidence in themselves \cite{williams2017individual}, especially their ability to detect phishing \cite{chen2020examination}, which can be improved via education and training \cite{moody2017phish}. This PF may correlate with too much self-confidence \cite{house2020phishing}.
In an experiment with 53 undergraduates students (34\% computer science majors, 66\% psychology majors), Hong et al. \cite{hong2013keeping} found that approximately 92\% of participants misclassified phishing emails even though 89\% had earlier indicated that they were confident of their ability to identify phishing emails.
            

\item {\sc Absentmindedness}. This PF describes the degree to which one's attention is diverted from a particular task. Reb et al. \cite{reb2015mindfulness} found that employees' absentmindedness is positively related to emotional exhaustion, which negatively affects jobs performance. Absentminded people can easily click phishing links because they do not pay attention to what they are doing \cite{zafar2019traditional,collier2020port}.

\end{enumerate}
It is intuitive that {\em cognitive} PFs play important roles in influencing individuals' susceptibility to social engineering attacks. However, our understanding is superficial.

\smallskip

\noindent{\bf Emotion PFs}.
These PFs describe human feelings, motivational states, and approaches or avoidance behaviors. They include so-called {\em visceral triggers}, which are strong internal drivers to satisfy a basic need.
\begin{enumerate}
\item {\sc Greed}. 
This PF is well recognized \cite{mondal2019review, kano2021trust, alyahya2021understanding, wang_2017, alabdan2020phishing, chiew2018survey, yasin2019contemplating} and
describes one's intense and selfish desire for something, especially wealth, power, or food. 
{\sc Greed} is one of the persuasion tools used in phishing attacks \cite{siadati2017mind} and is often paired with need
(i.e., the attacker knows what a victim needs and present what the victim needs as a bait) \cite{ferreira2018ransomware}.
Greed is recognized by some researchers as a human limitation when comparing human-based security versus technology-based security \cite{salahdine2019social, adil2020preventive}. Moreover, the greedier a person is, the more likely the person will fall victim to social engineering attacks \cite{li2019towards}.

\item {\sc Fear}. This PF, which tends to make people vulnerable \cite{aldawood2019reviewing, tu2019users, chiew2018survey, Jensen02}, describes one's belief that something painful, dangerous or threatening may happen. It is relevant because situations that evoke fear bring about a strong avoidance reaction in both behavioral responses and cognitive processing \cite{ness2017reactions}. Algarni et al.  \cite{algarni2017empirical} found that social engineering attacks are effective against people who submit to fear toward, and orders from, influential people. Siadati et al. \cite{siadati2017mind} treated fear as one of the phishing persuasion techniques.

\item {\sc Sympathy}.
This PF describes the emotional state of individuals who understand the mental or emotional state of another person without actually feeling the same emotion \cite{williams2017individual}. Sympathy is a universally virtue that has the potential risks of making people vulnerable to social engineering attacks \cite{wang2021social}. Indeed, attackers often seek to gain people's sympathy \cite{gallegos2017social, benias2018hacking,kano2021trust}. 


\item {\sc Empathy}. This PF, which is one of the human vulnerability that breaks a victim's defense \cite{arabia2020social, greavu2014social, schurmann2020effective,airehrour2018social},
describes the emotional state of an individual who personally relates to the mental or emotional state of another person based on their own experiences with the same state.
For example, scammers often exploit empathy as an intuitive behavior \cite{abe2019deploying} or a persuasion technique \cite{siadati2017mind} to get what they want from victims \cite{williams2017individual,abe2019deploying}.

\ignore{
\item {\sc Disgruntlement}. This factor describe one's anger or  dissatisfaction towards another. It can lead to 
insider threats in organizations \cite{bell2019insider,osuagwu2015mitigating} where disgruntled employees take revenge 
against organization policies or voluntarily expose organization to risks. The US Secret Service and Carnegie Mellon studied 49 cases of insider sabotage, and found that in 88\% of the cases, the perpetrator held a work-related grievance before the act of abuse \cite{willison2009motivations}.
}

\item {\sc Loneliness}. This PF describes one's subjective perception of discrepancy between the desired and the actual social companionship, connectedness, or intimacy \cite{buecker2020loneliness}. 
It is often exploited by social engineering attacks because the feeling of alienation from peers makes people vulnerable 
\cite{williams2017individual,lekati2018complexities}. The psychological reason is that attackers can exploit the need for attention that accompanies the feeling of loneliness, which may be even more relevant to elder people \cite{lin2019susceptibility}.
A study of 299 participants \cite{deutrom2021loneliness} found that loneliness positively predicts problematic internet uses that can be exploited by social engineering attacks. 
\end{enumerate}
The preceding discussion suggests that {\em emotion} PFs have been widely exploited by attacks. However, there is no deep or quantitative understanding of their impact on individuals' susceptibility to attacks.

\smallskip

\noindent{\bf Social Psychology PFs}.
These PFs describe one's interpersonal behaviors and often 
involve connection, influence, and demand/request interactions between the individual and one or more others.
There are 8 such PFs, among which the first 6 are derived from Cialdini's principles of persuasion.
\begin{enumerate}
\item {\sc Authority}. 
This PF describes power or dominance over someone \cite{aldawood975taxonomy}. 
Social engineering attackers use {\sc authority} to lure their victims to divulge confidential information, especially through spear phishing \cite{zheng2019session}. 
Bullée et al \cite{bullee2018anatomy} found that, out of the six principles of persuasion, the effect of authority alone exceeds that of the other five principles together.
It is worth mentioning that attackers often exploit {\sc authority} and {\sc scarcity} (which is described below) together \cite{zheng2019session}.
An empirical study with 612 participants
\cite{workman2007gaining} showed that individuals who are more obedient to authority succumb more frequently to social engineering attacks.

\item {\sc Reciprocation}. 
This PF describes the tendency to pay back a favor
done for them in the past \cite{lea2009psychology, lin2019susceptibility, ghafir2016social}. It is sometimes tied to people's urge to {\sc consistency} (described below) \cite{schaab2017social}. Bullée et al \cite{bullee2018anatomy} found that reciprocation is the third most used principle of persuasion exploited by social engineering attacks.

\item {\sc Liking (Similarity)}. 
This PF describes 
individuals' tendency to react positively to those with whom they hold some kind of relationship \cite{schaab2017social}. It reflects that people may be persuaded to obey others if they display certain favourable or familiar characteristics \cite{frauenstein2020susceptibility}. This PF has been exploited to create profiles that portray trusted traits or appear friendly to lure victims \cite{williams2017individual}. Bullee et al. \cite{bullee2018anatomy} found that {\sc liking} is widely exploited by social engineering attacks. Hatfield \cite{hatfield2018social} found that {\sc liking} is an individual variable that explains a person's tendency to fall victim to social engineer attacks.

\item {\sc Scarcity}. 
This PF describes the lack of goods/services and is used to lure their victims.
It has been widely exploited in online scams \cite{williams2017individual,bullee2018anatomy} and phishing emails.
For example, Heijden et al. \cite{van2019cognitive} found that {\sc scarcity} is a vulnerability trigger which social engineering attackers often craft in their phishing emails to push users to respond.
This PF is often exploited together with the {\sc authority} PF to lure victims into submitting to their demands \cite{zheng2019session,kearney2016can}.

\item {\sc Social Proof}. 
This PF describes one's tendency to imitate others regardless of the importance or correctness of the behavior \cite{algarni2017empirical, frauenstein2020susceptibility, van2019cognitive, moody2017phish, wang2021social}. 
It can put people at risk because they tend to let down their suspicion when everyone else appears to share the same or a similar behavior \cite{schaab2017social}.
In an experiment with 50,000 Facebook users,
Das et al. \cite{das2014increasing} found that users with ten or more Facebook friends tend to update their security settings after being informed that their friends have updated their own security settings. 

\item {\sc Consistency} (aka {\sc Commitment}).
This PF describes the degree to which one is dedicated to a person, object, task, or ideal \cite{wang2021social}. Social engineering attacks use commitment to persuade their victims \cite{ghafir2016social}. Algani et al. \cite{algarni2017empirical} found that dogmatic adherence to past decisions may influence the decisions a person will make in the future. Social engineering attacks can exploit this consistency to exploit victims without their knowledge \cite{van2019cognitive, kano2021trust}.

\item {\sc Disobedience}. This PF describes one's dogmatic refusal to obey authority or rules set forth by authority, which can make one 
susceptible to social engineering attacks \cite{collier2020port}. While it is well known that people who are more trusting and obedient to authority are more susceptible to social engineering attacks \cite{jampen2020don}, it is less know that willful disobedience of employees can also be exploited by social engineering attacks  \cite{kirlappos2014learning}.

\item {\sc Respect}. This PF describes an individual's esteem for another, which is the degree to which they are perceived as valuable or worthwhile to the individual in question \cite{algarni2017empirical}. For example, an individual may not question a suspicious request from a friend (e.g., an unsolicited email that contains a link) out of respect for their relationship \cite{redmiles2018acm}. This PF may be exploited together with the aforementioned {\sc authority} \cite{abe2019deploying, ghafir2016social}.

\ignore{the fear of losing one's job can 
overshadow reasoning in performing duties  \cite{katz2018report,hatfield2019virtuous}.\footnote{\color{red} This example demonstrates fear, not respect. It is also unclear from the description whether ``respect'' here refers to the individual's respect for another or the other's respect for the individual.}}
\end{enumerate}
The preceding discussion suggests the following PFs have been widely exploited by social engineering attacks: {\sc authority}, {\sc scarcity}, {\sc liking} ({\sc similarity}), and {\sc reciprocation}. Deep understanding of these PFs might shed light on the design of effective defense. For example, an effective defense may first identify whether an incoming email falls into the {\sc authority} category and if so tailored defenses may be used to decide whether the email is indeed from an authority; this would be more effective than using the same detector which treats all incoming emails equally without leveraging the PFs behind them.

\smallskip

\noindent{\bf Personality and Individual Difference PFs}.
These PFs are relatively stable and dispositional and differentiate one individual from another. For example, some people are habitually more meticulous and attentive to detail than others, while some people are habitually more trusting.
\begin{enumerate}
\item {\sc Disorganization}. This PF describes the tendency of an individual to act without prior planning or to allow their environment to become or remain unstructured or messy. These conditions may blind them to anomalies or cues of social engineering attacks, resulting in higher susceptibility \cite{collier2020port}.

\item {\sc Freewheeling}. This PF describes
the degree of one's disregard for rules or conventions and of their unconstraint or disinhibition. This PF contributes to ones' susceptibility to social engineering attacks \cite{collier2020port}.

\item {\sc Individual Indifference}. This PF describes the degree to which one shows disinterest toward an assigned or necessary task. A sustained indifference towards security can cultivate a culture of risky human behaviors, which can be exploited by social engineering attacks \cite{chowdhury2019impact}.

\item {\sc Negligence}. This PF describes an individuals' failure to take proper care during a particular task.
It is an important reason of security breaches \cite{safa2015information, adil2020preventive,li2019towards, ndibwile2019empirical}.
Li et al. \cite{li2019towards} reported that 27\% of data breaches are due to negligent employees or contractors, who usually have remote access to organizations' internal networks. 

\item {\sc Trust}. This PF describes the tendency of one to trust or believe in someone else (i.e., not doubting the honesty of others). People who are more trusting are more susceptible to social engineering attacks \cite{jampen2020don}, which is not surprising because developing trust is a key element of
social engineering attacks \cite{zheng2015entity, aldawood975taxonomy}. Moreover, people are predisposed to trust others they view as likable and phishers make use of this PF to scam victims \cite{hatfield2018social}. In a study with 612 participants,
Workman \cite{workman2007gaining} found that people who are more trusting succumb more frequently to social engineering attacks. 
\item {\sc 
Self Control}. This PF describes one's ability to regulate their decision-making processes in the face of strong emotions and desires. A lack of self control allows individuals to fall victim to online scammers \cite{williams2017individual}.
Individuals with low self-control tend to exhibit a higher willingness to take risks in 
situations that violate cybersecurity principles \cite{chowdhury2019impact, van2017big, holt2020testing}.
\item {\sc Vulnerability}. This PF describes the degree to which one is in need of special care, support, or protection because of age, disability, or risk of abuse or neglect.
In a study aiming to identify those at greater risk of falling victim to social engineering attacks in an organization, 
Bullee et al. \cite{bullee2017spear} find that employees with one year of service or less are more vulnerable to spear phishing (52.07\%) victimization compared to employees with eight years of services (23.19\%). 

\ignore{
\cite{chitrey2012comprehensive}
found that the most vulnerable groups of individuals are
new employees (41\% of participants), clients/customers (23\%), partners/contractors (12\%), top-level management (7\%), and IT professionals (17\%).\footnote{\color{red} These appear to be sample sizes. Which group is most vulnerable (or are they given in order? --- if so, specify that)? Or just cite that ``some'' are more vulnerable than others...}}

\item {\sc Impatience}. This PF describes one's frustration while waiting for a particular event to occur or at the length of time needed to accomplish a task. \cite{holt2020testing}. 
Impatient individuals may be more susceptible to social engineering attacks because they do not carefully examine contents or cues of social engineering attacks, especially when they focus on immediate gratification \cite{holt2020testing}.

\item {\sc Impulsivity}. This PF describes the tendency of one acting without much thought \cite{das2019sok}. In a study with 53 randomly selected participants (undergraduate students),
it was found that participants who scored low on impulsivity better managed phishing emails \cite{welk2015will}. In another study, it was found that individuals who are sensation-seeking, which is a form of impulsivity, were more likely to become scammed \cite{whitty2018you}.
            
\item {\sc Submissiveness}. 
This PF describes the degree of one's readiness to conform to authority or will of others. 
In a study with approximately 200 participants, it is found that 
high submissiveness implies a high susceptibility to phishing emails \cite{alseadoon2012more}.

            
            
\item {\sc Curiosity}. This PF describes the degree at which one desires to know something. Online scammers exploit victims' curiosity to encourage errors in judgement and decision-making \cite{williams2017individual} or serve as a persuasion technique to lure their victims \cite{siadati2017mind,xiangyu2017social}.
\item {\sc Laziness}. This PF describes the degree of one's voluntary inability to carry out a task with the energy required to accomplish it. 
Laziness makes people unwilling to do the necessary work or apply the effort to mitigate risk, and thus makes them more susceptible to
social engineering attacks \cite{wang2020into}. 

\item {\sc 
Vigilance}. 
This PF describes the degree that one is watchful for possible dangers or anomalies. A high vigilance makes one less vulnerable to social engineering attacks \cite{tu2019users,ndibwile2019empirical}. In a phishing experiment with 3000 university students, 
it was found that {\sc vigilance} reduced susceptibility to scams \cite{tu2019users}. However, even though an individual with a high {\sc vigilance}, who usually does not want to open a suspicious email, may actually end up opening it due to spontaneous {\sc curiosity} 
\cite{jalali2020employees}. This highlights the possible interactions between PFs, namely that one PFs may dominate another under certain circumstances, which explains the difficulty in coping with social engineering attacks.


\item {\sc Openness}.
This PF describes 
one's active imagination and insight \cite{cherry2012big}. Individuals with high openness are often curious about the world and other people, eager to learn new things, enjoy new experiences, and are more adventurous and creative.
High oppenness has been found to increase susceptibility to phishing attacks
\cite{frauenstein2020susceptibility,das2019sok}.

\item {\sc Conscientiousness}. This PF describes one's
thoughtfulness, impulse control, and goal-directed behaviors. People with high conscientiousness tend to be organized, mindful of details, self-disciplined, goal-oriented, proficient planners, and considerate about how their behaviors might affect others  \cite{cherry2012big,frauenstein2020susceptibility}.
It is found that people with a high conscientiousness are less susceptible to 
spear phishing attacks  \cite{halevi2015spear}.
            
\item {\sc Extraversion}. This PF, also known as {\sc extroversion}, describes the degree to which one is 
sociable, assertive, talkative, and 
emotionally expressive \cite{cherry2012big}. People with a high extraversion are outgoing and tend to gain energy in social situations. 
A study found that
{\sc extraversion} (and {\sc openness} and {\sc agreeableness}) increase one's susceptibility to 
phishing emails \cite{alseadoon2015influence}.
            
\item {\sc Agreeableness}. This PF describes one's attributes related to
trust, altruism, kindness, affection, and other prosocial behaviors \cite{cherry2012big}. 
A study found that people with a high
{\sc agreeableness} (and {\sc neuroticism}, which is described below) 
are more susceptible to phishing attacks \cite{yuan2019detecting}.

\item {\sc Neuroticism}. This PF 
describes one's moodiness and emotional instability.  People with high {\sc neuroticism} often exhibit mood swings, anxiety, irritability, and sadness \cite{cherry2012big}. 
Individuals with high
{\sc neuroticism} are more susceptible to phishing attacks \cite{yuan2019detecting}.
\end{enumerate}
We observe that enhancing some PFs (e.g., {\sc vigilance}) and reducing others (e.g., {\sc openness}) can reduce one's susceptibility to social engineering attacks. These should be leveraged to design future defenses. Moreover, the PFs are not independent of, or orthogonal to, each other. This suggests the importance of characterizing the relationships between them (e.g., ``{\sc openness} increases {\sc curiosity}'') because it would help identify the root cause of susceptibility to social engineering attacks.

\smallskip

\noindent{\bf Workplace PFs}.
These PFs have to do with the culture and organizational structure of workplace. This is relevant because various workplace environments may result in various levels of stress, employee engagement, or employee loyalty.

\begin{enumerate}
\item {\sc Workload}. This PF describes the amount of work that one has to do. A survey of 488 employees at three hospitals
showed that the level of employee workload is positively correlated with the likelihood of employees clicking on phishing links \cite{jalali2020employees}.
Another study found that subjective mental workload creates memory deficit that leads to an inability to distinguish between real and fake messages, increasing susceptibility to attacks \cite{aldawood2019reviewing}. 
            
\item {\sc Stress}. This PF describes 
the physical, emotional, or psychological strain on a person incurred by their environment. 
It has been found that when people are stressed, their ability to notice suspicious communications (e.g., distinguishing real from fake messages) is reduced, making them more susceptible to social engineering attacks \cite{williams2017individual,aldawood2019reviewing}. 

\item {\sc Busyness}. This PF describes the degree to which one has too much to do, which may or may not be associated with workload. 
People with a high {\sc busyness}
are more susceptible to 
phishing emails as they do not pay much attention to details \cite{chowdhury2019impact} or have reduced  
cognitive processing \cite{conway2017qualitative}. 

\item {\sc 
Hurry}. This PF describes the degree one is rushing to complete a task. Hurried people 
may not adhere to secure practices
because they reduces the amount of time available for the individual's active task \cite{chowdhury2019impact}; these people are susceptible to social engineering attacks under these circumstances
\cite{rastenis2020mail}.
            
\item {\sc Affective Commitment}. This PF describes one's emotional attachment to an organization. A study with 612 participants found that
people with a high {\sc affective commitment} more likely fall victim to social engineering attacks  \cite{workman2007gaining}.
            
\item {\sc Habituation}. This PF describes one's tendency to perform a particular task repeatedly. 
A study on how users perceive and respond to security messages using eye-tracking with 62 participants
found that people gazed less at warnings over successive viewings (i.e., they were more habituated to the warnings) and thus were less attentive to security warnings \cite{brinton2016users}. In other words, increased habituation increases susceptibility to attacks.

\end{enumerate}
The preceding discussion suggests that workplace PFs have a significant impact on individuals' susceptibility to social engineering attacks and should be taken into consideration when designing future defenses.

\subsection{Systematizing Psychological Techniques (PTs)}

Our analysis of the literature prompts us to consider the following 13 PTs.

\begin{enumerate}
\item {\em Urgency}. Urgency has an impact on cybersecurity when a victim is confronted with a situation which requires immediate action or is ostensibly under time pressure \cite{chowdhury2019impact}, such as decreasing the chance of detecting deceptive elements in a message \cite{vishwanath2011_dssystems}. It leverages the {\sc cognitive miser}, {\sc fear} and {\sc negligence} PFs. It is often used in scareware attacks to urge users to install software to avoid threats (e.g., viruses) or missing a plug-in which prevents them from viewing some desired contents \cite{nelms2016usenix}.   
    
\item {\em Attention Grabbing}. This technique uses visual and auditory elements to prompt a victim to focus attention on deceptive attack elements to increase compliance. It leverages the {\sc absentmindedness} and {\sc curiosity} PFs. The malvertising, scareware, and click-baiting attacks use attention grabbing along with visceral triggers and incentives (below) to encourage compliance \cite{nelms2016usenix}.

\item {\em Visual Deception}. This technique repurposes benign visual elements to induce {\sc trust} \cite{vishwanath2011_dssystems}. It leverages the {\sc overconfidence}, {\sc trust}, and {\sc habituation} PFs. The typosquatting and clone-phishing attacks exploit this technique by creating URLs that are visually similar to benign URLs. 
    
\item {\em Incentive and Motivator}. This technique encourages a desired behavior or compliance with a request. Incentive provides external rewards for action, while motivator provides internal rewards (i.e., gratification) for an individual. In social engineering attacks, incentive often leverages  visceral triggers, which are 
commonly used in malvertising and click-baiting attacks as well as in the Nigerian scam \cite{herley2012weis}. Motivator exploits {\sc sympathy}, {\sc empathy}, {\sc loneliness}, and {\sc disobedient}. Wire transfer scams
exploit victims' sympathy for the attacker as a motivator to encourage someone to transfer money to an attacker who claims to have made an erroneous money transfer. 

\item {\em Persuasion}. This technique encourages a particular behavior by exploiting the {\sc liking}, {\sc reciprocation}, {\sc social proof}, {\sc consistency}, and {\sc authority} PFs. The effectiveness of each persuasion technique depends on other things like age \cite{lin2019susceptibility} and request type \cite{goel2018mobile, alohali2018identifying}.  The use of persuasion is prevalent in email-based attacks such as phishing \cite{goel2018mobile,ferreira2015analysis, wright2014research}. 

\item {\em Quid-Pro-Quo}. Quid-Pro-Quo in Latin means "something for something". This technique attempts to make a victim willing to take risk on exchange for a high payoff (e.g., money, free services or avoiding embarrassment). It leverages the {\sc reciprocation}, {\sc greed}, and {\sc dishonesty} PFs \cite{stajano2011understanding}. For example, an attacker can impersonate a police officer to make a victim pay for illegal content (e.g., pornography) on the victim's computer \cite{heartfield2015taxonomy}; otherwise, the attacker threatens with arresting the victim for the possession of illegal content. In the Nigerian Prince Scam (419) \cite{herley2012weis}, the Quid-Pro-Quo is the expectation that the victim give a small amount of money to receive a larger amount of money later. 
    
\item {\em Foot-in-the-Door}. This technique attains compliance for a large request by making small requests over time \cite{freedman1966compliance}. It exploits the {\sc consistency} PF. It is commonly used in honey trap and catfishing.
    
\item {\em Trusted Relationship}. This technique exploits an existing trust relationship by taking advantage of the {\sc authority}, {\sc respect,} and {\sc trust} PFs. 
For example, through LinkedIn (a trusted service provider), an attacker posing as a recruiter can connect to employment-seeking victims \cite{allodi2019need}; spamdexing (SEO) exploits a user's trust in a search engine provider's (e.g., Google) results; Business Email Compromise exploits the trusted relationship between an executive staff officer and a subordinate employee. 

\item {\em Impersonation}. This technique assumes a false identity to increase a victim's compliance. It exploits the {\sc authority}, {\sc respect} and {\sc trust} PFs. In OSN-based attacks like Honey Trap, an attacker uses fake profiles to lure victims into interacting with them  \cite{algarni2017empirical}; an attacker using Business Email Compromise assumes the persona of a senior executive to exploit their authority by prompting a victim to transfer money to an account \cite{junger2020fraud}. 
 
\item {\em Contextualization}. This technique projects an attacker as a member of the victim's group in order to establish commonality with potential victims and increase the success of attacks \cite{goel2017got,rajivan2018creative}. 
It is often used in attacks like whaling, 
catfishing, and drive-by downloads \cite{goel2018mobile}. 
    
\item {\em Pretexting}. This technique increases the engagement of victim with the attacker.
It leverages the {\sc trust} PF. For example, phishing emails can use this technique to increase responsiveness by adding elements that refer to current events like holiday festivities or news \cite{alhamar2010ieee, goel2017got}. 
    
\item {\em Personalization}. This technique uses personal information to tailor messages or express similar interest to the victim to engender trust \cite{hirsh2012psysci, jagatic2007acm}. It exploits the {\sc personality} and {\sc individual differences} PFs. 
    
\item {\em Affection trust}. This technique establishes an affectionate relationship with a victim. It exploits the {\sc affective commitment} PF. Affection does not lower risk perceptions or increase trust, but makes an individual more willing to take risks and thus increases compliance \cite{mcallister1995affect}.  It is commonly used in catfishing and honey traps. 
\end{enumerate}
As we will see, these PTs help build bridges to map social engineering attacks and the PFs they exploit.




\section{Systematizing Social Engineering Attacks}
\label{sec:attacks}

This section presents the objectives of social engineering attacks and a taxonomy of them. This would help us understand how attackers may choose specific attacks based on their objectives, which could help us design effective defenses against threats of given objectives. It is worth mentioning that in order to see if we overlooked some literature investigating these attack names, we conducted another round of search by using the attack names as keyword in the digital libraries 
mentioned above. This leads to 6 papers which were not identified in the previous search. In addition, we were aware of two social engineering attacks which are discussed in online materials but not academic literature, namely {\em Honey Trap} \cite{copado-2021} and {\em Angler Phishing} \cite{fraudwatch-2017,experian-2018}, which we included as well. 

\subsection{Attack Objectives}

We categorize social engineering attacks according to the following four main types of attack objectives.
\begin{enumerate}
\item {\em Getting access to systems}.
Social engineering attacks are often used as a first step of full-fledged attacks against networked systems (e.g., advanced persistent threats).

 \item {\em Stealing money}. Social engineering attacks such as phishing are often used to steal victims' money.


\item {\em Stealing sensitive information}. Social engineering attacks such as phishing are often used to steal sensitive information such as passwords.


\item {\em Revenging}. Social engineering attacks can be used to take revenge against enterprises, organizations, or individuals by 
releasing damaging information about them
\cite{chitrey2012comprehensive}.


\end{enumerate}


\subsection{Attacks} 

Figure \ref{fig:SocialEngineering} highlights the taxonomy of Internet-based social engineering attacks based on the medium they leverage: email vs. website vs. online social network (OSN). It is worth metioning that these attacks relate to each other; for example, the below-mentioned {\em drive-by download} attack may leverage various kinds of phishing emails to deceive a victim to visit malicious websites. These attacks are elaborated below.



\begin{figure}[htbp!]
\centering
\includegraphics[width=.8\textwidth]{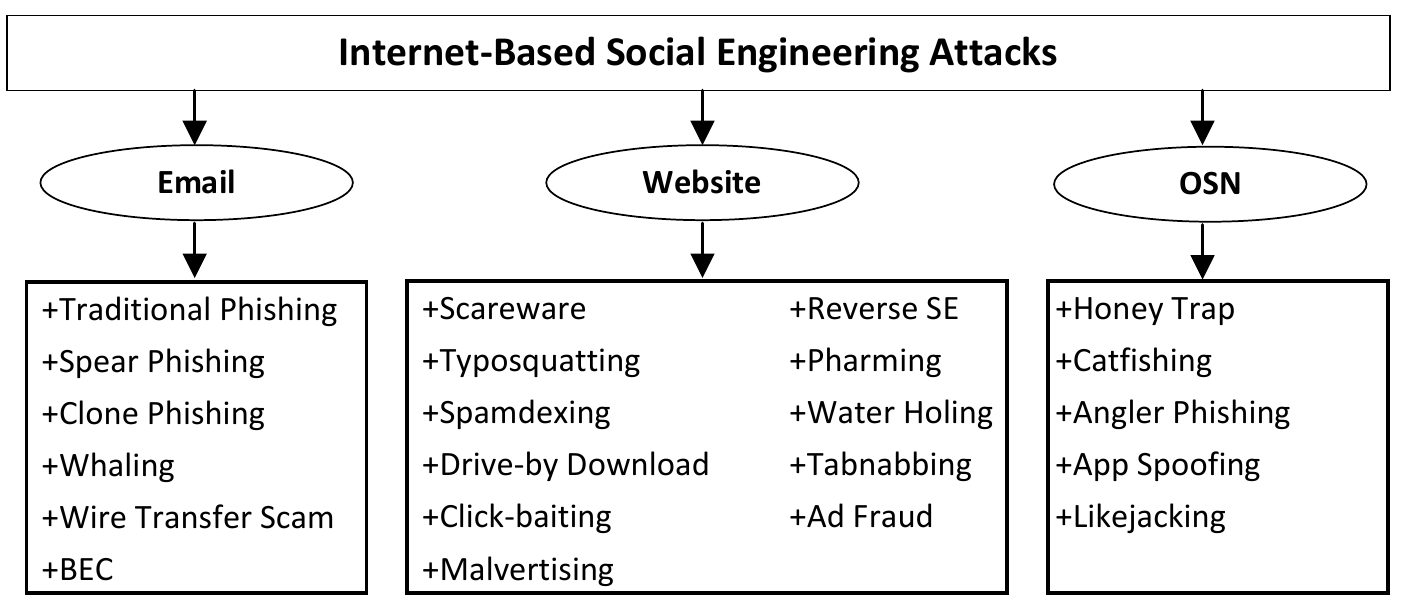}
\caption{Taxonomy of social engineering attacks exploiting emails, websites, and online social networks (OSN), where BEC stands for Business Email Compromise.}  
\label{fig:SocialEngineering}
\end{figure}

\ignore{
\begin{figure}[htbp!]
\centering
\includegraphics[scale=1.0]{images/InternetAttacks2.pdf}
\caption{Taxonomy of social engineering (SE) attacks exploiting emails, websites, and online social networks (OSN). BEC = Business Email Compromise. {\color{red}the box on the top is wasting too much space ... also, can you make these pictures into pdf to make them higher resolution like Figure 1? no reason to use such fuzzing pictures}} 
\label{fig:SocialEngineering}
\end{figure}
}

\noindent{\bf Email-based Attacks}.
This category includes six attack techniques, which are varying flavors of phishing. These attacks are largely complementary to each other.
\begin{enumerate}
\item {\em Traditional Phishing}. In this attack, a phishing email is sent without a particular target in mind, but with the hope that someone will fall victim to it (i.e., no personalization in such phishing emails) \cite{ho2019detecting, dou2017systematization, steves2019phish}. 
This attack is often motivated to steal money.
This attack exploits the {\sc greed} factor because it attempts to entice victims for rewards such as in the 419 scam that promises a large amount of money if a victim pays a small amount of money.

\item {\em Spear Phishing}. A spear phishing email contains information personalized for a specific target,
usually addressing the target by name and title.
This attack is often motivated to steal money, get access to systems, steal sensitive information, or for revenge. This attack exploits the {\sc authority} factor because it attempts to deceive a victim into believing that the phisher/attacker is a person of authority and a victim must act promptly \cite{ho2017detecting, heartfield2015taxonomy, goel2018mobile}.

\item {\em Clone Phishing}. Such an email is cloned from a previously sent/received email, replaces its links and/or attachments with malicious ones, and spoofs the legitimate sender's email address so that the target would not not suspect the clone email \cite{bhavsar2018study, alam2020phishing, prem2019phishing}. 
This attack is often motivated to steal money and sensitive information. This attack exploits the {\sc trust} factor because it attempts to make a victim think that the cloned email is a continuation of a previous communication and in order to help comply with the attacker's request.     

\item {\em Whaling}. A whaling email is similar to a spear phishing email by targeting specific individuals. Unlike spear phishing which can target arbitrary individuals, whaling emails target management, such as CEOs \cite{goel2018mobile, heartfield2015taxonomy, salahdine2019social}. This attack is often motivated to steal money, get access to systems, steal sensitive information, or for revenge.
This attack exploits the {\sc trust} factor because it attempts to deceive, for example, a CEO into believing in the content an email and then following the instructions described in it, often by impersonating someone that the victim knows.

\item {\em Wire Transfer Scam}.
In this attack, an email is sent to targeted individuals in order to deceive the individual into sending money (via, for example, Western Union) to pay for services or goods \cite{burch2015wire}.  The attacker often impersonates a service company, such as a utility, that threatens that the victim's services will be cut off immediately unless a wire transfer is made, ans sometimes impersonate reputable individuals \cite{chaganti2021recent}.
It is motivated to steal money and exploits the {\sc fear} factor because it threatens to cut services to victims.

\item {\em Business Email Compromise (BEC)}. 
This attack 
uses email frauds against private, government and non-profit organizations,
by targeting specific employees with spoofed emails impersonating a senior colleague, such as the CEO or a trusted customer \cite{cidon2019high, venkatesha2021social}. 
This attack is motivated by the objective of stealing money \cite{cidon2019high}. It exploits the {\sc trust} factor because it attempts to deceive victims into thinking that they are paying a legitimate bill for goods/services received from a trusted party.
\end{enumerate}

\noindent{\bf Website-based Attacks}.
This category includes 11 attacks. These attacks are not necessarily complementary or orthogonal to each other because one attack may leverage another as a supporting technique (e.g., {\em Ad Fraud} may use {\em malvertizing} as a support technique).

\begin{enumerate}

\item {\em Scareware}. This attack is to pop up a window with warning content which tells the user that the computer has been infected by malware and that the user should click a link or call a number shown on the pop-up window to get help. The attacker's intent is to scare the user to click the link or to call the number shown on the pop-up window, which will give the attacker the opportunity to access the user's sensitive information or ask the user to send a gift card number to have the problem fixed remotely. 
Most scareware do not harm the computer, but are instead used to scare victims to provide information or money to the attacker \cite{or2019dynamic}. This attack exploits the {\sc fear} factor because it scares victims into thinking that their computer is compromised and needs immediate attention.
    
\item {\em Typosquatting} (or {\em URL Spoofing)}. This attack takes a user to a malicious website when the user mistypes a character in a URL,
such as mistyping {\tt www.bank0famerica.com} for {\tt www.bankofamerica.com}, where the former mimics the latter in terms of website content while incorporating a malicious payload \cite{heartfield2015taxonomy}.
This attack exploits the {\sc negligence} factor because it anticipates individuals mistyping.
    
\item {\em Spamdexing} (or {\em Search Engine Poisoning}). This attack tricks a search engine to list a malicious website on the top of the list returned by a search \cite{heartfield2015taxonomy}.
It is effective because many users trust the search results listed on the top and treat them as most relevant, causing them to most likely visit them.
It exploits the {\sc trust} factor because it anticipates that users treat the websites on the top of search results as most relevant.

\item {\em Drive-by Download}. This attack is used to compromise a vulnerable browser when it visits a malicious or compromised website, possibly prompted by phishing emails containing the malicious URL \cite{ProvosHotbot07}.
It exploits the {\sc trust} factor because a victim may trust the website in question, or the {\sc vulnerability} factor when a victim is not aware of this attack, or the {\sc negligence} factor when a user does not update/patch a browser or does not pay careful attention to recognize malicious websites.

\item {\em Click-baiting}. This attack is to place an enticing text/image on a web page to draw the attention of visitors so that they click on a link to a malicious or compromised website \cite{meinert2018really}.
An example is a message on a website reading "Betty reveals how she gets to 100 years of age without ever doing sports". 
It exploits the {\sc curiosity} factor because it entices victims to click on the link to figure out more information.

\item {\em Malvertising}. This attack abuses advertisement 
such that when a user clicks on the advertisement, the user may be redirected to a malicious website 
\cite{chiew2018survey}. 
It exploits the {\sc trust} factor because victims think they are getting legitimate ads.  It also takes advantage of the {\sc negligence} factor when victims do not perform due diligence.

\item {\em Reverse Social Engineering}. This attack
creates a situation causing a victim to contact the attacker \cite{irani2011reverse}.
It exploits the {\sc trust} factor because it puts a victim in a situation of need, thereby contacting the attacker.

\item {\em Pharming}. 
This attack builds malicious websites to steal money or sensitive information from victims when visiting them \cite{adil2020preventive}.
It exploits the {\sc trust} factor because victims do not think these websites are malicious and the {\sc negligence} factor because victims do not perform due diligence.

\item {\em Water Holing} This attack exploits vulnerabilities of third party websites to attack victims when visiting them \cite{wang2021social}. 
This attack if often waged to steal money or sensitive information. 
It exploits the {\sc trust} factor because victims trust that the websites that they are visiting to be secure.

\item {\em Tabnabbing}. This attack attempts to deceive a victim into visiting a malicious website which mimics a legitimate website and asks the victim to login into the malicious website,  while making the victim think that the malicious website is the legitimate website and forwarding the victim's login credential to the legitimate website \cite{salahdine2019social}.
It often leverages the same origin policy of browsers, where a second page on a browser can access scripts from another page as long as both pages have the same origin \cite{steffens2019don}.
It attempts to steal sensitive information (e.g., login credentials).
It exploits the {\sc absentmindedness} factor because a victim thinks that a previously visited website is asking for login credentials again.

\item {\em Ad Fraud}. 
This attack exploits ads to defraud advertisements, where the fraudster deceives the victims that are using a platform to advertise their goods and services by generating fake traffic (possibly via malvertising, scareware, click-baiting, and likejacking) \cite{kanei2020detecting}.
It often attempts to steal money in the sense that the ads do not incur real traffic from real users, but forged traffic instead.
It exploits the {\sc trust} factor because victims believe that they are getting legitimate traffic to their advertisements.
\end{enumerate}

\noindent{\bf Online Social Network-based (OSN-based) Attacks}. This category includes five attack techniques.
    
\begin{enumerate}
\item {\em Honey Trap}. This attack targets a particular victim with a love-related relationship and may be seen as the counterpart of spear phishing. 
For example, John knows that Philip likes blonds and thus creates a fake profile of a blond on Instagram to like and comment on Philip's posts; Philip sees a blond liking his posts and thinks it is an opportunity for him to meet a blond; once a relationship is established, John can deceive Philip in many ways, including financial extortion \cite{copado-2021}. 
Our evaluation shows that this attack exploits the {\sc loneliness} factor because lonely people turn to the platform to seek attention.

\item {\em Catfishing}.
This attack creates a fake persona to seek online dating to lure victims interested in the persona, similar to the traditional phishing because the attack does not target a specific victim \cite{simmons2020catfishing}.
For example, the attacker posts as women to lure men to send them money for made-up reasons, for example, ``My Internet service will be suspended for accumulated bills, please help me pay or I'll not be able to chat with you if my Internet is suspended". This attack exploits the {\sc liking} ({\sc similarity}) factor because victims have the tendency to react positively to someone that they have some relationship with \cite{schaab2017social}. 

\item {\em Angler Phishing}.
This attack is used to lurk among the comments posted by users on social forums, like yelp, and then takes advantage of any comment that may need a resolution to \cite{fraudwatch-2017}. For example, an attacker may see a comment of a customer complaining about a bank transaction or a purchase. The attacker then poses as a customer satisfaction specialist of that company and asks the customer for detailed information in order to address the customer's problem \cite{experian-2018}. An unsuspecting customer may give away personal information with the hopes that the problem will be resolved, not knowing that they have been phished.  This attack exploits the {\sc vulnerability} factor because frustrated victims desperately need solutions and the {\sc trust} factor that victims put in the service companies.

\item {\em App Spoofing}. This attack uses bogus apps to spoof legitimate ones on platforms which are less regulated than (for example) iPhone App stores or Google Play Store. 
When a user uses the same credential for multiple platforms, the attacker can steal a user's credentials to get access to the user's account on other platforms \cite{malisa2017detecting}. 
It exploits the {\sc openness} and {\sc curiosity} factors because users who are open and curious will often try new things.

\item {\em Likejacking}. This is the social media version of click-jacking attack. This attack places a transparent layer (e.g. transparent iframe) on a legitimate webpage so that when a user clicks anywhere on the webpage, the user is actually clicking on the transparent layer which directs the user to the attacker's website 
\cite{alabdan2020phishing,calzavara2020tale}.
In Likejacking, when a user sees the ``like'' button on a Facebook post, on top of which there is a transparent layer not visible to the user, the user may click on the page and then be directed to a malicious website. 
This attack exploits the {\sc liking and similarity} factor because the attacker sets the trap knowing that people tend to like comments of people they follow on OSN.
\end{enumerate}

\subsection{Discussion} 

Attackers have been exploiting PFs to wage website-based social engineering attacks.
For example, a phishing email can be crafted to exploit PFs like {\sc fear}, {\sc authority}, and {\sc curiosity}, causing victims to react in a manner desired by the attacker. In principle, exploiting PFs increases the likelihood that a victim will overlook important cues of attacks.

\begin{insight}
Social engineering attackers have made due effort at identifying and exploiting the relevant human PFs for waging attacks.
\end{insight}

\section{Systematizing Social Engineering Defenses}
\label{sec:defenses}

Similar to the systematization of attacks, we naturally divide defenses into three categories: email-based, website-based, and online social network-based attacks. Although it is intuitive to present defenses with respect to each attack mentioned above, this is less constructive because one defense may be able to defend against multiple attacks. Since our systematization is centered on PFs, we further divide defenses into two sub-categories: those that do not leverage PFs and those that leverage PFs. This makes it easy to recognize which PFs have been leveraged for defense purposes.

\subsection{Defenses against Email-based Attacks}

\noindent{\bf Defenses Not Leveraging PFs}. Most studies on defenses against email-based social engineering attacks fall into this category.
Defenses against various kinds of phishing have been extensively investigated, for which we refer to previous surveys for a large body of literature \cite{khonji2013phishing, ali2020survey, jain2021survey, sahu2014survey, alabdan2020phishing}.
Ho et al. \cite{ho2017detecting} proposed using anomaly detection to identify 
real-time credential spear phishing attacks in enterprise settings.
Ho et al. \cite{ho2019detecting} proposed a classifier for detecting lateral phishing emails, which are spear phishing emails that are sent by an attacker after compromising some computers in a network and are seemingly coming from colleagues. 
Cidon et al. \cite{cidon2019high} proposed a defense against Business Email Compromise attacks by leveraging
supervised learning and statistics about email histories.
All these defenses, including those which are surveyed in the previous literature, leverage technological aspects (e.g., statistical data).


\noindent{\bf Defenses Leveraging PFs}. There are few studies falling into this category. These primarily focus on {\em eye tracking}, which is related to the {\sc vigilance} PF. One study 
\cite{pfeffel2019user} leverages eye tracking to investigate gaze patterns when individuals are reading phishing emails. However, it showed: (i) even in the best-case scenario, when individuals are expecting phishing emails and are motivated to discover them, many cannot distinguish a legitimate email from a phishing email; and (ii) well-crafted phishing emails can still fool 40\% of the participants. This means that leveraging eye tracking is not effective. 
Nevertheless, another study \cite{heartfield2018detecting} shows that incorporating a human in the defense loop can substantially reduce the success rate of some spear phishing attacks from 81\% to below 10\%. This highlights the importance of incorporating humans into the defense, but it is not clear how the participants exactly achieved this and whether this can be generally applied to other settings.


\subsection{Defenses against Website-based Attacks}
There are many studies on detecting malicious websites, including website-based social engineering attacks. The simplest method would be to use blacklists to filter malicious websites. However, the trustworthiness of blacklists is questionable because they may be outdated and/or be provided in a black-box fashion without justification \cite{XuCODASPY13}.

\smallskip

\noindent{\bf Defenses Not Leveraging PFs}.
There are many studies in this sub-category, primarily leveraging Artificial Intelligence/Machine Learning (AI/ML).
For example, VisualPhishNet 
\cite{abdelnabi2020visualphishnet} leverages visual similarities between websites to detect phishing websites;
Phishpedia 
\cite{linphishpedia} leverages deep learning to detect phishing websites via identity logos;
Mnemosyne 
\cite{allen2020mnemosyne} is a postmortem forensic analysis engine  for accurately reconstructing, investigating, and assessing the ramifications of watering hole attacks;
Mao et al. \cite{mao2018detecting} 
presents a method to detect phishing websites by leveraging learning-based aggregation analysis to decide page layout similarity;
Nakamura and Dabashi \cite{nakamura2019proactive} propose to detect new phishing sites by leveraging 
domain name generation and other attributes.
    
Another approach is to leverage hardware features.
For example, 
Fidelius \cite{eskandarian2019fidelius} is an architecture which uses trusted hardware enclaves to protect sensitive user information from potentially compromised browsers and operating systems;
FIDO (First IDentity Online) is a web-authentication mechanism for mitigating phishing attacks in real time, by leveraging one-time-password as a second factor for authentication \cite{ulqinaku2020real}.

\smallskip

\noindent{\bf Defenses Leveraging PFs}.
There are few studies falling into this sub-category. One of them is 
\cite{aladawy2018persuaded}, which presents a game-based 
training against social engineering attacks by leveraging 
social psychology PFs with the help of cards. The game is designed to provide knowledge and train people through social psychology theories on resistance to persuasion. 
This game is further enhanced 
to contain more content and 
to accommodate contexts \cite{goeke2019protect}.

\subsection{Defenses against Online Social Network-based Attacks}
There are many studies on detecting OSN-based social engineering attacks. Since defenses against website-based social engineering attacks can be leveraged to defend against some OSN-based social engineering attacks, we will focus on the defenses that are unique to OSN-based attacks.

\smallskip

\noindent{\bf Defenses Not Leveraging PFs}. 
These defenses primilarily leverage AI/ML techniques. 
For example, Yuan et al. \cite{yuan2019detecting} present a method to Sybil accounts;
Xu et al. \cite{xu2021deep} present 
a method to detect abusive accounts in OSNs;
Wang et al. \cite{wang2020into} present a 
chatbot to actively collect intelligence 
to help detect e-commerce frauds.

\smallskip

\noindent{\bf Defenses Leveraging PFs}.
The only study we are aware of that falls into this sub-category is 
\cite{junger2017priming}, which investigates the effectiveness of two defense interventions:
one is to prime the user with cues to raise awareness about the dangers of social engineering attacks and the other is to warn against the disclosure of personal information. They find that warnings do help improve the user's behavior but most users do not adjust their behaviors when monetary rewards are at stake. This does suggest the importance of incorporating PFs into defenses.

\subsection{Discussion} 
The preceding section suggests that current defenses primarily leverage technological solutions, especially AI/ML, but rarely incorporates PFs; when a defense does incorporate a human in the loop, a much higher effectiveness can be expected \cite{heartfield2018detecting}. Since social engineering attacks primarily exploit weaknesses in human information processing, we argue that effective defenses would have to incorporate the ``right'' PFs because they are the ``root cause'' of the problem in a sense, where the ``right'' factors need to be precisely pinned down in future studies. 

\begin{insight}
Current defenses against social engineering attacks have achieved limited success because they do not adequately take into account human PFs.
\end{insight}


\section{Systematizing the Relationships between Attacks, PTs, PFs, and Defenses}
\label{sec:sok}

\label{ss.mapping}


\begin{figure*}[!htbp]
\centering
\includegraphics[width=\textwidth,keepaspectratio]{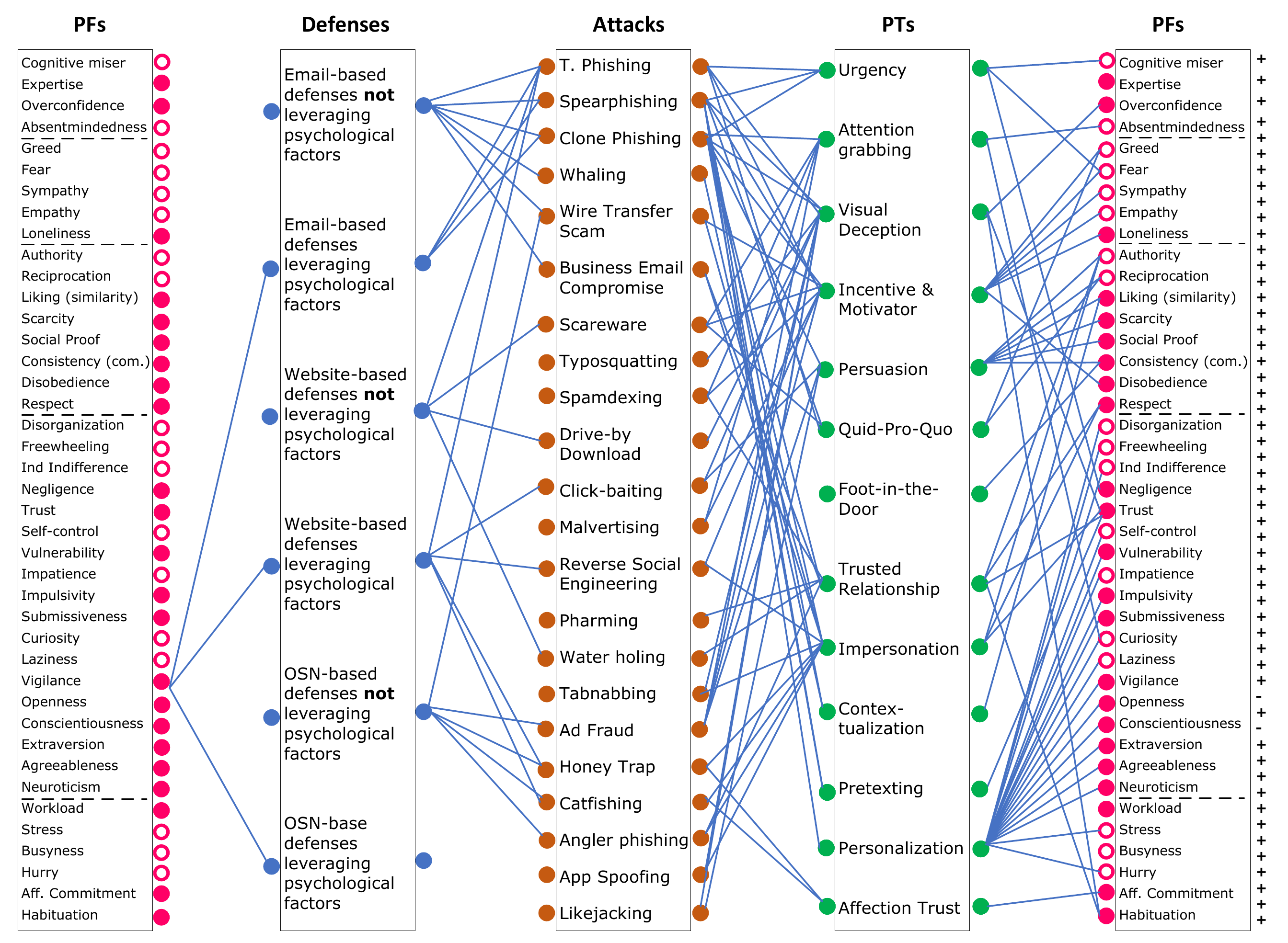}
\caption{Relationships between the PFs, the defenses, the attacks, and the PTs, where the PFs are separated by dashed lines into the aforementioned five categories described in the text. The PFs with empirical quantitative studies are represented with a filled circle and a circle otherwise. The ``+'' (``-'') sign indicates that a factors increases (decreases) susceptibility; Ind = individual, com = commitment, T = Traditional. 
}
\label{fig:mapping}
\end{figure*}


\ignore{dropped from the initial table: Persuasion-FITD,}

Figure \ref{fig:mapping} systematizes the aforementioned relationships.

\smallskip

\noindent{\bf Relationships between Defenses and Attacks}.
First, {\em email-based defenses not leveraging PFs} have been proposed to defend against the following attacks: Traditional  Phishing, Spear phishing, Clone Phishing, Whaling, Wire Transfer, and BEC (Business Email Compromise). 
Whereas, {\em email-based defenses leveraging PFs} have been proposed to defend against Traditional Phishing, Spear Phishing, and Clone Phishing.
Second, {\em website-based defenses not leveraging PFs} have been proposed to defend against Traditional Phishing, Scareware, Drive-by Download, and Watering Holes.
Whereas, {\em website-based defenses leveraging PFs} have been proposed to defend against Traditional Phishing, Click-baiting, Reverse Social Engineering, Honey Trap, and Catfishing.
Third, {\em OSN-based defenses not leveraging PFs} have been proposed to defend against 
OSN-based Honey Trap, Catfishing, and Angler phishing.
Whereas, we are not aware of any {\em OSN-based defenses leveraging PFs}.

\ignore{
\begin{enumerate}
\item Human Sensor: Wire Transfer Scam
\item AI/ML: wire transfer scam, Ransomware, Business Email Compromise, Ad Fraud, Scareware Malvertising
\item Dongle: Pharming, Tabnabbing
\item Blacklisting: Traditional Phishing, Spear Phishing, Clone Phishing, Ransomware, Ad Fraud
\item New Technologies: Ransomware, Pharming, Water Holing, Ad Fraud
\item Eye Tracking: Phishing
\item Best Practices: Clone-Phishing, Typosquating, scareware,
\end{enumerate}}

\smallskip

\noindent{\bf Relationships between Attacks and PTs}. Social engineering attacks leverage PTs to take effect on PFs. In what follows we systematize the PTs leveraged by email-based, website-basedm and OSN-based attacks, respectively.


First, PTs leveraged by email-based social engineering attacks are summarized as follows.
(1) Traditional Phishing leverages Urgency, Visual Deception, Incentive and Motivator, and Quid-Pro-Quo.
(2) Spear Phishing leverages Urgency, Visual Deception, Incentive and Motivator, Quid-Pro-Quo, Contextualization, Pretexting, and Personalization.
(3) Clone Phishing leverages Urgency, Attention Grabbing, Visual Deception, Incentive and Motivator, Persuasion,  Trusted Relationship, Impersonation, Pretexting, and Personalization.
(4) Whaling leverages the Contextualization technique. 
(5) Wire Transfer Scam leverages Incentive and Motivator as well as Impersonation.
(6) BEC (Business Email Compromise) leverages Trusted Relationships and Impersonation. 

Second, PTs leveraged by website-based attacks are summarized as follows.
(1) {Scareware} leverages Quid-Pro-Quo, Incentive and Motivator, and Attention Grabbing. 
(2) {Typosquatting} leverages Visual Deception. 
(3) {Spamdexing} leverages Trusted Relationship, Incentive and Motivator, and Attention Grabbing. 
(4) {Drive-by Download} leverages Visual Deception. 
(5) {Click-Baiting} leverages Persuasion and Visual Deception. 
(6) {Malvertising} leverages Incentive and Motivator, and Attention Grabbing. 
(7) {Reverse Social Engineering} leverages Incentive and Motivator as well as Impersonation.
(8) {Pharming} leverages Trusted Relationship. 
(9) {Water Holing} leverages Trusted Relationship. 
(10) {Tabnabbing} leverages Visual Deception and Impersonation. 
(11) {Ad Fraud} leverages Persuasion, Incentive and Motivator, Attention Grabbing, and Visual Deception.

Third, PTs leveraged by OSN-based attacks are summarized as follows.
(1) {Honey Trap} leverages  Impersonation and Affection Trust. 
(2) {Catfishing} leverages Impersonation and Affection Trust. 
(3) {Angler Phishing} leverages Impersonation and Trusting. 
(4) {App Spoofing} leverages Impersonation and Visual Deception.
(5) {Likejacking} leverages Persuasion and Visual Deception. 

\smallskip

\noindent{\bf Relationships between PTs and PFs}. Although a PT can exploit multiple PFs, we observe that a given PT often exploits PFs within a single psychological category. This prompts us to systematize their relationships according to the five categories of PFs as follows.
First, PTs exploiting {\em cognition PFs} are summarized as follows.
(1) {Attention Grabbing} exploits the {\sc absentmindedness} and {\sc curiosity} factors.
(2) {Visual Deception} exploits {\sc overconfidence}, {\sc trust}, and {\sc habituation}.
Second, PTs exploiting {\em emotion PFs} are summarized as follows. 
(1) {Urgency} leverages {\sc cognitive miser}, {\sc fear} and {\sc negligence}.
(2) {Incentive and Motivator} leverages {\sc greed}, {\sc fear}, {\sc sympathy, empathy, loneliness}, and {\sc disobedient}.
Third, PTs exploiting {\em social PFs} are summarized as follows.
(1) {Persuasion} leverages the {\sc liking, reciprocation, social proof, consistency,} and {\sc authority} factors.
(2) {Quid-Pro-Quo} leverages {\sc Reciprocation}, and {\sc Greed}.
(3) {Foot-in-the-Door} leverages {\sc Consistency}.
(4) {Trusted Relationship} leverages {\sc authority}, {\sc respect,} and {\sc trust}.
(5) {Impersonation} leverages {\sc authority}, {\sc respect} (i.e., close relationship) and {\sc trust}.
(6) {Contextualization} leverages {\sc liking} ({\sc similarity}).
Fourth, PTs exploiting {\em individual differences and personality PFs} are summarized as follows.
(1) {Pretexting} leverages the {\sc trust} factor.
(2) {Personalization} leverages {\sc disorganized}, {\sc freewheeling}, {\sc individual indifference}, {\sc negligence}, {\sc trust}, {\sc self control}, {\sc vulnerability}, {\sc impatience}, {\sc impulsivity}, {\sc submissiveness}, {\sc curiosity},{\sc laziness}, {\sc vigilance}, {\sc openness}, {\sc conscientiousness}, {\sc extraversion}, {\sc agreeableness} and {\sc neuroticism}.
Fifth, PTs exploiting {\em workplace PFs} are summarized as follows. The {\em Affection Trust} leverages  {\sc affective commitment}.

\smallskip

\noindent{\bf Relationships between Defenses and PFs}. To our knowledge, {\sc vigilance} is the only PF that has been considered in defenses against email-based, website-based and OSN-based attacks \cite{tu2019users, aldawood975taxonomy, aldawood2019reviewing, alsharnouby2015phishing, junger2020fraud} . This means that much research is needed before leveraging PFs to design effective defenses. This also prompts us to explore future research directions in the next section.

\ignore{Warnings help direct attention and trigger . In contrast, training increases vigilance by affecting risk perceptions. The main relationships include the following mappings.
\begin{enumerate}
    \item Warnings. This defense leverages the following factors: Vigilance
    \item Training:. This defense leverages the following factors: Vigilance
\end{enumerate}}

\ignore{\footnote{revise the rest of Section VI.A by mimicking what I did above: consistency!!!!!!!!! the terms must be exactly the same as what they are described above; also, the order of PTs in this case (and factors in the next) MUST BE the same as they are introduced so that it is for a reader to check!!!!!! Both are consistency issues!!!!!!!!!!}}
\ignore{\footnote{{\color{red}fit the content into the preceding structure}}}

\smallskip

\noindent{\bf Summary}. Figure \ref{fig:mapping} leads to the following. First, PTs are widely exploited for attacks but are rarely incorporated into defenses. This discrepancy gives attackers a big advantage. Second, some PTs are leveraged by attacks more often than others. For example, impersonation, attention grabbing and visual deception are widely used across email-based, website-based, and OSN-based social engineering attacks. This means that future defenses should target such PTs. Third, there is a good potential to leverage workplace PFs to design effective defenses because they cannot be manipulated by attackers. 

\begin{insight}
\label{insight:answer-to-why}
Current defenses have achieved a limited success as they rarely incorporate human PFs. 
\end{insight}


\section{Future Research Directions}
\label{sec:research-directions}

The preceding systematization suggests that effective defenses should be guided by psychological principles. This prompts us to seek psychological principles that can guide the design of effective defenses.

\smallskip

\noindent{\bf Guiding Principle: The Theory of System 1 vs. System 2 in Human Information Processing}. The human mind processes information from the environment likely through a variety of cognitive mechanisms. 
A popular theory is centered at the distinction of System 1 (heuristic) vs. System 2 (analytic)  \cite{kahneman2011thinking}. System 1 is fast, effortless, based on heuristics, and often thought of as error-prone; whereas, System 2 is slow, effortful, and involves deep analytical thinking \cite{kahneman2011thinking}. Putting another way, the theory suggests that deliberate reasoning, which is typically logical or mathematical, falls into the scope of System 2 \cite{doi:10.1177/0963721419855658}. 
While common in psychology, the theory does not come without criticism, especially the fact that despite the characteristics of the two processes is often clear, but the factors that determine when an individual will think analytically or rely on their intuition is unclear \cite{pennycook2015makes}.
This has inspired research in 
cognitive psychology \cite{lin2022thinking}
using electroencephalography (EEG) to decipher their respective underlying neural mechanisms \cite{williams2019thinking}, and in improving Artificial Intelligence learning from human decision making capabilities \cite{booch2020thinking}. 
There have also been attempts to improve this dual-thinking process to a three-stage dual process model of analytic engagement \cite{pennycook2015makes}. 
Moreover, a recent development is that humans can actually process deliberate reasoning involving logical principles in an intuitive fashion (i.e., without deliberation) 
\cite{doi:10.1177/0963721419855658}. This sheds light on leveraging the theory to guide us in understanding social engineering attacks through the psychological lens and in designing new defenses without forcing humans to trap into System 2 when dealing with social engineering attacks.

The preceding systematization and guiding principle prompt us to propose the following research directions: (i) use psychological principles to guide the design of a {\em qualitative} framework to describe social engineering attacks; (ii) conduct empirical studies to {\em quantify} the effect of PFs on human susceptibility to social engineering attacks; and (iii) leverage these quantitative findings to guide the design of effective defenses.

\subsection{Creating a Qualitative Psychological Framework Tailored to Social Engineering Attacks} 

Our premise is that in the context of social engineering attacks, both heuristics and analytic processing can help prevent victimization under different conditions, and it may not be accurate to regard heuristics as uniquely error-prone. We thus use this framework as a starting point towards a more robust framework later in the paper.

In order to design effective defenses, we need to understand how the human information processing procedure interprets information associated with these attacks. 
This prompts us to propose a framework for describing human information processing of materials associated with social engineering attacks.

\begin{figure}[!htpb]
\centering
\vspace{-1em}
\includegraphics[scale=0.4]{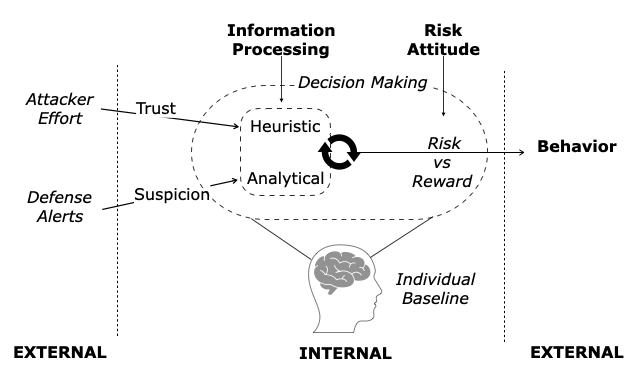}
\caption{A qualitative psychological framework describing human information processing of social engineering attacks.}
\label{fig:unified_framew}
\end{figure}

Figure \ref{fig:unified_framew} highlights the framework we propose. It is inspired by the state-of-the-art System 1 vs. System 2 framework. The new framework has two internal components, {\em information processing} and {\em risk attitude}, which collectively determine one's {\em behavior} (e.g., clicking a link in a phishing email or not) based on: 
(i) {\em individual baseline}, namely the PFs that may be influenced by an attack to benefit the attacker;
(ii) the external {\em attacker effort} at earning one's trust (e.g., how real a phishing email looks); and (iii) {\em defense alerts} (e.g., one's competence in raising suspicions against phishing emails). The framework is elaborated below. 

\smallskip

\noindent{\bf Risk Attitude}.
We propose incorporating risk attitude because studies show that
it affects the likelihood of social engineering victimization and it may be independent of human information processing \cite{conway2017qualitative,lea2009psychology}. This is not surprising because risk attitude affects motivators which drive humans to act \cite{maslow1943theory}. 
There are three well-known risk attitudes: {\em risk seeking}, {\em risk aversion} and {\em risk-neutral}.
For example, even when information processing triggers suspicions, a risk-seeking user may still comply with a malicious request because the prospect of reward exceeds the perceived risk, explaining why some people make risky decisions in cyberspace especially when they feel they have less to lose \cite{conway2017qualitative, howe2012ieee} and why some people still fall victim even if they recognize the risk \cite{lea2009psychology}. 

\smallskip

\noindent{\bf Heuristic vs. Analytic Processing}. This is inspired by the System 1 vs. System 2 framework.
However, our goal is to identify the conditions that would push one into {\em heuristic processing} or {\em analytic processing}, respectively. 
\begin{itemize}
\item \emph{Heuristic processing}. This uses patterns and rules, or a trial-and-error approach, to reach a decision. Heuristics are often used in situations of uncertainty where information or time is limited. In social engineering attacks, non-experts are more likely to rely on heuristics to determine the credibility of the message. 
In addition to rules and patterns, non-expert users also rely on previous experiences (often acquired through trial and error) to detect social engineering messages \cite{abbasi2016ieee, redmiles2018acm}. Heuristics are useful, but can cause errors when the rules used to determine credibility are based on elements that can be manipulated by attackers, or when they are unable to discriminate between benign and social engineering messages. 

\item \emph{Analytic processing}. This involves evaluating multiple factors to reach a decision. It requires that an individual is knowledgeable on factors relevant to the outcome and have the information required to support the decision. In social engineering attacks, individuals with cybersecurity {\sc expertise} are more likely to use analytic processing to detect social engineering messages. For example, experts would consider multiple factors to determine the credibility of emails
and attend to suspicious elements in them \cite{kumaraguru2006acm}.
\end{itemize}
The preceding categorization is important because attackers often attempt to deceive victims into heuristic processing but not analytic processing to increase their chance of success.
It remains to be investigated how heuristic processing and analytic processing would work together.

\smallskip

\noindent{\bf Individual Baseline}.
These are the PFs that affect information and risk attitude. 
As discussed above, the following PFs encourage the use of heuristic processing: {\sc habituation}, {\sc stress} and {\sc workload}. {\sc habituation} because they reduce suspicions.
Moreover, a combination of {\sc stress} and {\sc workload} increases the reliance on heuristic processing and decreases {\sc vigilance}.

\smallskip

\noindent{\bf Attacker Effort}.
This is an attacker's effort at exploiting human PFs to earn victims' trust and encourage their compliance. 
For example, an attacker can earn trust from a victim by creating emails of high quality and appealing to the victim. 
This is because many users judge credibility based on superficial attributes, like the professional appearance of a website, absence of grammatical errors, or recognizable logos in emails \cite{dhamija2006acm, kim2005dq}. 
To generate an appealing message, an attacker can exploit a combination of PTs (e.g., persuasion, personalization, contextualization). Having identified the PTs exploited by attacks, it is an important open problem to investigate {\em how} attacks influence PFs, which in turn influences the heuristic vs. analytic processing as mentioned above.


\smallskip

\noindent{\bf Defense Alerts}. These include the mechanisms that are employed to warn users of potential threats to trigger their {\sc vigilance}. Intuitively, an effective alert would cause users to switch their attention to the warning information and maintain their attention long enough to process it. An effective warning would trigger suspicion \cite{montanez2020human}, such as cue salience triggering attention switching \cite{wogalter2018communication}. It is an important open problem to investigate
how defenses can influence PFs to offset the influence of attackers.

Future research needs to confirm, disputes, or refine the qualitative framework.

\subsection{Creating a Quantitative Psychological Framework Tailored to Social Engineering Attacks}
The preceding qualitative framework, or it refined version, paves a way for quantifying the effectiveness of social engineering attacks and defenses. Specifically, we propose a hierarchical quantitative framework to describe individuals' {\tt susceptibility} to social engineering attacks as 
\begin{eqnarray*}
\text{{\tt susceptibility}}
=
f(\text{{\tt processing\_route}}, ~\text{{\tt risk\_attitude}}),
\label{eq:SIndex}
\end{eqnarray*}
where $f$ is a family of mathematical functions that are to be identified by future studies, {\tt processing\_route} means the use of heuristic or analytic processing, and {\tt risk\_attitude} indicates how the individual trades risk for reward. Moreover, the {\tt processing\_route} would be determined by

\begin{eqnarray*}
\text{{\tt processing\_route}}=g(\text{{individual\_baseline}}, 
~\text{{attacker\_effort}},~\text{{defense\_alerts}}).
\label{eq:ip-route}
\end{eqnarray*}
where $g$ is another family of mathematical functions that are to be determined by future studies, and {individual\_baseline}, {attacker\_effort} and {defense\_alerts} are defined above.

Future research needs to enrich the quantitative framework, or its refinement, with characteristics of the impact of the variables, including the relevant PFs and PTs. We envision that the resulting quantitative framework will be seamlessly incorporated into broader frameworks for investigating cybersecurity from a holistic perspective, such as Cybersecurity Dynamics \cite{XuCybersecurityDynamicsHotSoS2014,XuBookChapterCD2019,xu2020cybersecurity} which is driven by the need to quantify cybersecurity from a holistic perspective \cite{XuSciSec2021SARR,Pendleton16,XuSTRAM2018ACMCSUR}. Indeed, human susceptibility to social engineering attacks has been explicitly described in the mathematical models of preventive and reactive cybersecurity dynamics \cite{XuTDSC2011,XuTAAS2012,XuTDSC2012,XuIEEETNSE2018,XuIEEEACMToN2019,XuTNSE2021-GlobalAttractivity}. Moreover, human susceptibility to social engineering attacks needs to be adequately incorporated into other kinds of models, such as adaptive, proactive, and active cyber defense dynamics \cite{XuTAAS2014,XuQuantitativeSecurityHotSoS2014,XuHotSOS14-MTD,XuInternetMath2015ACD,XuHotSoS2015}. Since human susceptibility to social engineering attacks would not be independent from individual to individual as shown above (e.g., people exhibiting similar PFs would be susceptible to social engineering attacks to a similar extent), this kind of dependence needs to be explicitly considered in holistic cybersecurity models, as highlighted in \cite{XuInternetMath2012,XuQuantitativeSecurityHotSoS2014,XuInternetMath2015Dependence,XuGameSec13,XuHotSoS2018Firewall,XuTDSC2021DynamicDivresity}.

\smallskip



\subsection{Using Psychological Principles and Quantitative Findings to Guide Design of Effective Defenses} 

The qualitative framework discussed above suggests approaches to designing effective defenses as follows. First, under the premise that the standard theory of System 1 vs. System 2 is perfectly suitable for describing the phenomena resulting from social engineering attacks, effective defense should strive to train users in enhancing their System 1 decision-making when coping with potential social engineering attacks. Moreover, if the aforementioned recent development --- humans can actually process deliberate reasoning involving logical principles in an intuitive fashion (i.e., without deliberation) 
\cite{doi:10.1177/0963721419855658} --- is true, then this insight can be leveraged to build effective defense. These psychological principles play a fundamental role because it is not feasible to force humans to use System 2 when dealing with social engineering attacks simply because of (for example) the large amount of emails they will process on a daily basis.

The quantitative understanding resulting from the quantitative framework mentioned above would offer insights into designing effective defenses. The basic idea is to identify the important factors, namely the most relevant PFs and PTs, such that defenses can be tailored to influence them to minimize individuals' {\tt susceptibility} to social engineering attacks.
For example, if {\sc trust} turns out to be an important factor, then defenses can be tailored to minimize individuals' {\sc trust} (e.g., making everyone practice zero-trust on everything coming from Internet may be an effective defense against social engineering attacks).
As another example, if warnings can reduce individuals' {\tt susceptibility}, then it is important to investigate how to make warnings effective (e.g., using dynamic warnings instead of static warnings in order to reduce {\sc habituation} \cite{brinton2016users}).

\subsection{Summary}

The discussion presented above reflects our firm belief that effective defenses cannot be achieved without the guidance of sound psychological principles that are tailored to the domain or social engineering attacks. 

\begin{insight}
A quantitative psychological theory tailored to the social engineering domain is needed to guide the design of effective defenses.
\end{insight}


\section{Conclusion}
\label{sec:conclusion}

In order to understand why current defenses against social engineering attacks have achieved limited success, 
we have systematized human PFs and PTs (psychological factors and techniques) which have been exploited by attackers in particularly crafty ways. Our systematization of these attacks and current defenses against them highlights a key discrepancy which can explain the limited success of current defenses: defenses do not consider human PFs to the same degree that attacks do. This prompts us to propose a systematic roadmap for future research towards effective defenses.

\smallskip

\noindent{\bf Acknowledgement}. We thank Eric Ficke and Shawn Emery for feedback and proofreading the paper.

\ignore{

\section{Appendices}

If your work needs an appendix, add it before the
``\verb|\end{document}|'' command at the conclusion of your source
document.

Start the appendix with the ``\verb|appendix|'' command:
\begin{verbatim}
  \appendix
\end{verbatim}
and note that in the appendix, sections are lettered, not
numbered. This document has two appendices, demonstrating the section
and subsection identification method.

\ignore{
\section{SIGCHI Extended Abstracts}

The ``\verb|sigchi-a|'' template style (available only in \LaTeX\ and
not in Word) produces a landscape-orientation formatted article, with
a wide left margin. Three environments are available for use with the
``\verb|sigchi-a|'' template style, and produce formatted output in
the margin:
\begin{itemize}
\item {\verb|sidebar|}:  Place formatted text in the margin.
\item {\verb|marginfigure|}: Place a figure in the margin.
\item {\verb|margintable|}: Place a table in the margin.
\end{itemize}
}
\begin{acks}
To Robert, for the bagels and explaining CMYK and color spaces.
\end{acks}
}
\bibliographystyle{ACM-Reference-Format}
\bibliography{paper,rosa-ref,metrics1_2}


\end{document}